\documentclass[10pt,conference]{IEEEtran}
\IEEEoverridecommandlockouts
\def\BibTeX{{\rm B\kern-.05em{\sc i\kern-.025em b}\kern-.08em
    T\kern-.1667em\lower.7ex\hbox{E}\kern-.125emX}}
\ifCLASSINFOpdf
\fi
\usepackage{verbatim}
\usepackage{algorithm}
\usepackage{graphicx}
\usepackage{bm}
\usepackage{amssymb,amsfonts}
\usepackage{algorithmicx}
\usepackage[noend]{algpseudocode}
\usepackage{setspace}
\usepackage{float}
\usepackage{subfigure}
\usepackage{epstopdf}
\usepackage{amsthm}
\newtheorem{lemma}{Lemma}

\usepackage{stfloats}
\usepackage{enumerate}
\usepackage{multirow}
\usepackage{latexsym}
\usepackage{multicol,lipsum}
\usepackage{cuted} 

\usepackage{tikz}
\usepackage{textcomp}
\usepackage{xcolor}
\usepackage{cancel}

\usepackage{cite}

\algdef{SE}[DOWHILE]{Do}{doWhile}{\algorithmicdo}[1]{\algorithmicwhile\ #1}%
\usepackage{amsmath}
\allowdisplaybreaks
\makeatletter 
\def\@eqnnum{{\normalsize \normalcolor (\theequation)}}

\makeatother 
\begin{document}
\title{Resource Allocation for Multi-waveguide Pinching Antenna-assisted Broadcast Networks
}
\author{ Ruotong Zhao,~\IEEEmembership{Student Member,~IEEE,} Shaokang Hu,~\IEEEmembership{Member,~IEEE,} Deepak Mishra,~\IEEEmembership{Senior Member,~IEEE,}\\ and Derrick Wing Kwan Ng,~\IEEEmembership{Fellow,~IEEE} 
\\
\thanks{D. W. K. Ng and R. Zhao are supported by the Australian Research Council's Discovery Projects DP240101019 and DP230100603, respectively.}
\thanks{D. Mishra is supported by the Australian Research Council's Discovery Early Career Researcher Award (DECRA) DE230101391.}
\IEEEauthorblockA{School of Electrical Engineering and Telecommunications,  University of New South Wales, Sydney, NSW 2052, Australia
}}

\maketitle


\begin{abstract}
In this paper, we investigate the resource allocation for multi-dielectric waveguide-assisted broadcast systems, where each waveguide employs multiple pinching antennas (PAs), aiming to maximize the minimum achievable rate among multiple users. To capture realistic propagation effects, we propose a novel generalized frequency-dependent power attenuation model for dielectric waveguides PA systems. 
We jointly optimize waveguide beamforming, PA power ratio allocation, and antenna positions via a block coordinate descent scheme that capitalizes on majorization–minimization and penalty methods, circumventing the inherent non-convexity of the formulated optimization problem and obtaining a computationally efficient sub-optimal solution. Simulation results demonstrate that our proposed framework substantially outperforms both conventional antenna systems and single-PA-per-waveguide configurations, clearly illustrating the intricate trade-offs between waveguide propagation loss, path loss, and resource allocation among multiple PAs.
\end{abstract}
\large\normalsize
\section{Introduction}
To fulfill the rapidly growing data traffic demands anticipated in the upcoming sixth-generation (6G) wireless networks, multiple-input multiple-output (MIMO) technology incorporating flexible antenna systems has been widely envisioned as a critical enabler for effectively exploiting the extra spatial degrees of freedom (DoF) by dynamically reshaping wireless channels~\cite{wu2023movable}.
Depending on their specific implementations, a flexible antenna can be implemented by leveraging mechanical drivers~\cite{ma2024movable, ma2023compressed} or liquid metal~\cite{wong2020fluid}, with the latter known as fluid antennas; however, their repositioning range is typically limited to only a few wavelengths, often rendering them insufficient to maintain reliable line-of-sight (LoS) connectivity~\cite{ding2025flexible}. 
Considering that emerging communication systems are shifting toward higher frequencies, pinching antenna (PA) technology has emerged as a novel flexible antenna solution by utilizing dielectric waveguides activated by selectively pinching different dielectric particles to simultaneously regulate path loss and phase~\cite{xu2025pinching}. Such an approach provides stronger LoS connectivity and lower path loss compared with alternative flexible antennas~\cite{ding2025flexible}.
Moreover, the authors in~\cite{wang2025modeling} developed a physics-based hardware model, showing that PAs' power can be actively controlled by adjusting their coupling lengths along the waveguide.

Despite various efforts to unlock the full capabilities of PAs, prior studies have often overlooked the intrinsic characteristics of waveguides, instead simply assuming a uniform power distribution among multiple PAs~\cite{xu2025pinching, ding2025flexible, shan2025exploiting}. 
While the impact of dielectric waveguide propagation was examined in~\cite{hu2025, shan2025multigroup}, the former study was limited to single PA per waveguide scenarios, whereas the propagation model for the latter one overlooked the frequency impact and only depended on the location of the PA. These existing simplistic models significantly limit resource allocation accuracy, particularly at higher frequencies~\cite{balanis2024balanis}.
Although the authors in~\cite{shan2025exploiting} highlighted that PA systems are notably suitable for broadcasting scenarios, they neglected power allocation and realistic propagation conditions that severely impact performance. 
On the other hand, optimizing the positions of multi-PAs in multi-waveguide systems is especially challenging due to the coupled effects of path loss and phase alignment. Previous approaches, either simplifying the positioning problem into antenna activation selection~\cite{xu2025pinching} or employing sequential positioning for individual PAs while fixing the configuration of others~\cite{shan2025exploiting}, have underutilized the advantages of joint optimization, thus compromising both performance and scalability.

Motivated by these identified research gaps and associated practical challenges, this paper investigates resource allocation design for a multi-PA-assisted broadcast system employing a generalized practical waveguide power distribution model. Within the proposed framework, we jointly optimize the beamforming vectors across the waveguides, individual power ratios for each PA, and their spatial positions, explicitly accounting for the intertwined impacts of path loss and phase alignment.
The key contributions of our work are as follows: $\textbf{i)}$ We propose a realistic PA-based broadcasting system model that explicitly considers dielectric waveguide propagation involving multiple PAs within multiple waveguides. To facilitate system optimization, we establish a model to characterize the relationship between PA positions and their corresponding power attenuation in dielectric waveguides; $\textbf{ii)}$ We formulate the design as a resource allocation optimization problem. 
Its intrinsic non-convexity, arising from tightly coupled optimization variables, is handled by an iterative block coordinate descent (BCD) framework that alternately optimizes beamforming, PA power ratio allocation, and PA positions.
Due to the persistent non-convexity of each subproblem, beamforming and PA power ratios are updated by exploiting majorization–minimization (MM), whereas a penalty-augmented MM scheme tackles the amplitude-coupled phase terms in PA positioning. $\textbf{iii)}$ Our numerical results demonstrate that employing a multi-PA-per-waveguide system significantly improves the performance of broadcasting networks. The proposed design effectively exploits the additional DoF by adaptively optimizing the PAs' positions and power ratio, which is particularly beneficial in high-frequency scenarios.

\textbf{Notation:} 
Bold uppercase/lowercase letters denote matrices/vectors. $\mathbf{A}^T$, $\mathbf{A}^H$, $\text{Tr}(\mathbf{A})$, $\text{Rank}(\mathbf{A})$, and $[\textbf{A}]_{i,j}$ represent the transpose, Hermitian transpose, trace, rank, and the $(i,j)$-entry of the matrix $\textbf{A}$, respectively. $\mathbf{A} \succeq \mathbf{0}$ indicates a positive semi-definite matrix; $\text{diag}(\textbf{a})$ represents a diagonal matrix from the vector $\textbf{a}$, $\text{blkdiag}\left(\textbf{A}_1, ..., \textbf{A}_n\right)$ denotes a block diagonal matrix with $\textbf{A}_1, ..., \textbf{A}_n$, and  $\text{Diag}(\textbf{A})$ is a vector extracted from
the main diagonal elements of the matrix $\textbf{A}$. $\mathbb{C}^{N \times M}$, $\mathbb{R}^{N \times M}$, and $\mathbb{H}^N$ denote the set of all $N \times M$ matrices with complex and real entries, and all $N \times N$ Hermitian matrices, respectively. The largest eigenvalue of matrix $\textbf{A}$ and its associated eigenvector are denoted by $\Phi_{\rm max}\hspace{-1mm}\left(\hspace{-0.2mm}\textbf{A}\hspace{-0.2mm}\right)$ and $\bm{\Phi}_{\rm max}\hspace{-1mm}\left(\hspace{-0.2mm}\textbf{A}\hspace{-0.2mm}\right)$, respectively.
$|\cdot|$, $\left(\cdot\right)^*$, $\mathbb{R}^+$, and $\mathbb{E}\{\cdot\}$ denote the absolute and the conjugate value, positive real numbers, and the statistical expectation, respectively. Moreover, $\|\cdot\|$, $\|\cdot\|_2$, and $\|\cdot\|_*$ denote the norm, nuclear norm, and spectral norm, respectively. We use $\text{vec} \{\cdot\}$ to represent vectorization; $j = \sqrt{-1}$, $\Re(\cdot)$ and $\Im(\cdot)$ indicate the real and imaginary parts of a complex number, respectively. The circularly symmetric complex Gaussian (CSCG) distribution is denoted by $\mathcal{CN}(\mu, \sigma^2)$, with mean $\mu$ and variance $\sigma^2$, and $\sim$ stands for ``distributed as".

\section{System Model}\label{sec:System}
\begin{figure}[t]
\centering
\includegraphics[width=3.4in]{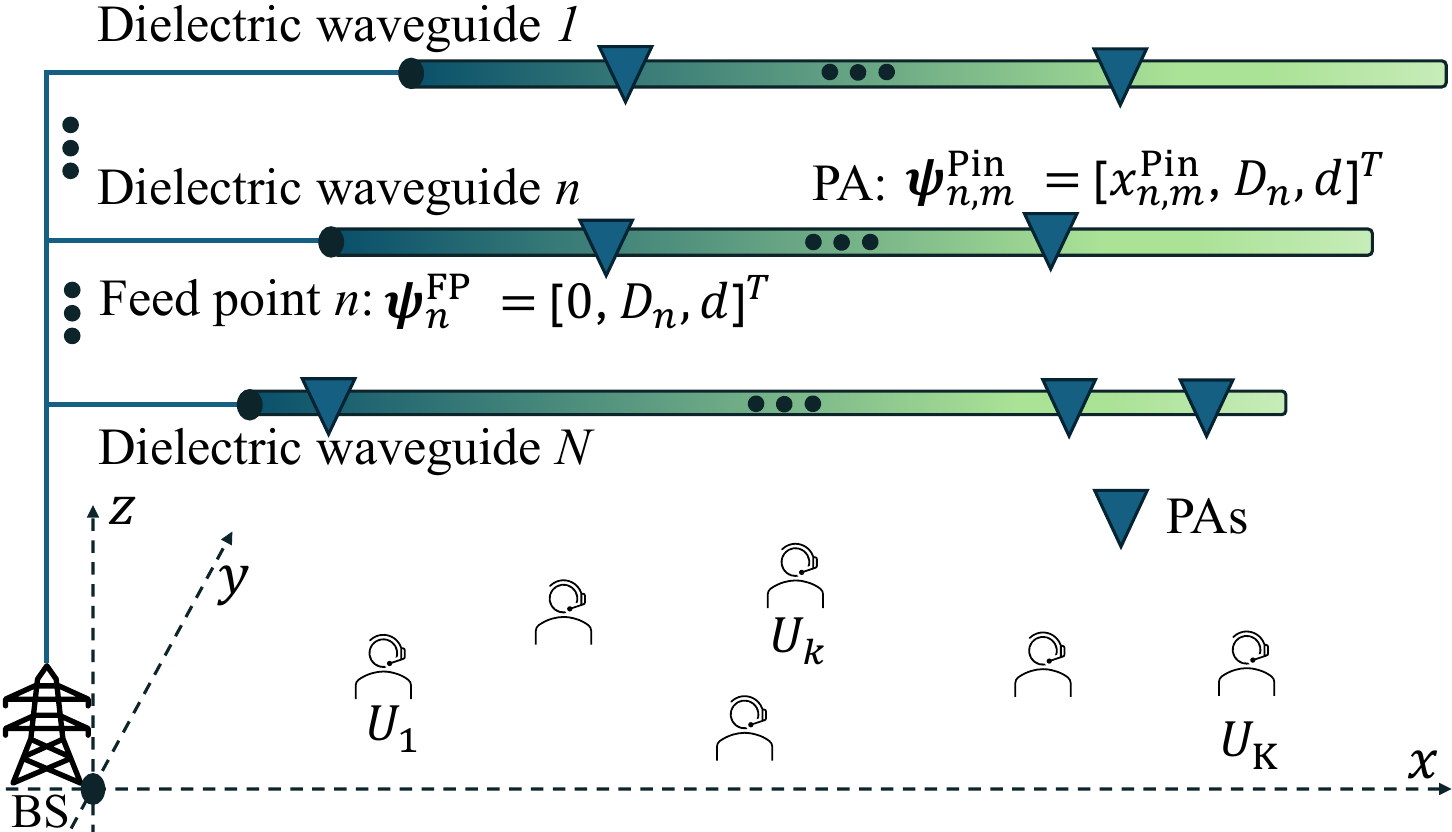}
\caption{A downlink PA-assisted communication system.}
\label{fig:SYSTEM}
\vspace{-0.5em}
\end{figure}
As illustrated in Fig. 1, we consider a downlink PA-assisted broadcast system, where a base station (BS) feeds $N$ dielectric waveguides, each with a length of $L$ meters and mounted at a height of $d$ meters. We assume that each waveguide accommodates $M$ PAs\footnote{One possible implementation of the PA system may involve mounting PAs on movable platforms along pre-installed tracks parallel to the waveguides~\cite{ding2025flexible}.
} and denote the waveguide set as $\mathcal{N} \triangleq \{1, \dots, N\}$. In the proposed scenario, all waveguides carry an identical transmitted signal $s \in \mathbb{C}$ with $\mathbb{E}\{|s|^2\} = 1$, simultaneously serving $K$ users who request the same content\footnote{The extension to multigroup multicasting is left for future investigation.}, e.g., live public events streaming and mass notifications, where the set of users is denoted as $\mathcal{K} \triangleq \{1, \dots, K\}$. 
We adopt a three-dimensional (3D) Cartesian coordinate system to depict the locations of critical components in our system model. Specifically, signals from the BS are fed into each waveguide at designated feed points and the beamforming precoder employed at the BS is denoted as $\textbf{w} \in \mathbb{C}^{N \times 1}$ with elements $w_n$, $|w_n|^2$ is the input power for the $n$-th waveguide, $\forall n \in \mathcal{N}$. In particular, the feed point of the $n$-th waveguide is denoted by $\bm{\psi}_n^{\mathrm{FP}} = \left[0,D_n,d\right]^T$, where $D_n$ represents the coordinate along the $y$-axis for the $n$-th waveguide. The position of the $m$-th PA on the $n$-th waveguide is expressed as $\bm{\psi}_{n,m}^{\mathrm{Pin}} = \left[x^{\mathrm{Pin}}_{n,m}, D_n, d\right]^T$, $\forall n \in \mathcal{N}, m \in \{1, \dots, M\}$, where $x^{\mathrm{Pin}}_{n,m}$ denotes the coordinate along the $x$-axis for the respective PA. We denote $u_{k}$ as the $k$-th user and its 3D position is written as $\bm{\phi}_k = \left[x_k, y_k, 0\right]^T$, $\forall k \in \mathcal{K}$, with coordinates $x_k$ and $y_k$ corresponding to the user's location along the $x$-axis and $y$-axis, respectively.

\subsection{Waveguide Propagation Model}
\begin{figure}[t]
\centerline{\includegraphics[width=3.4in]{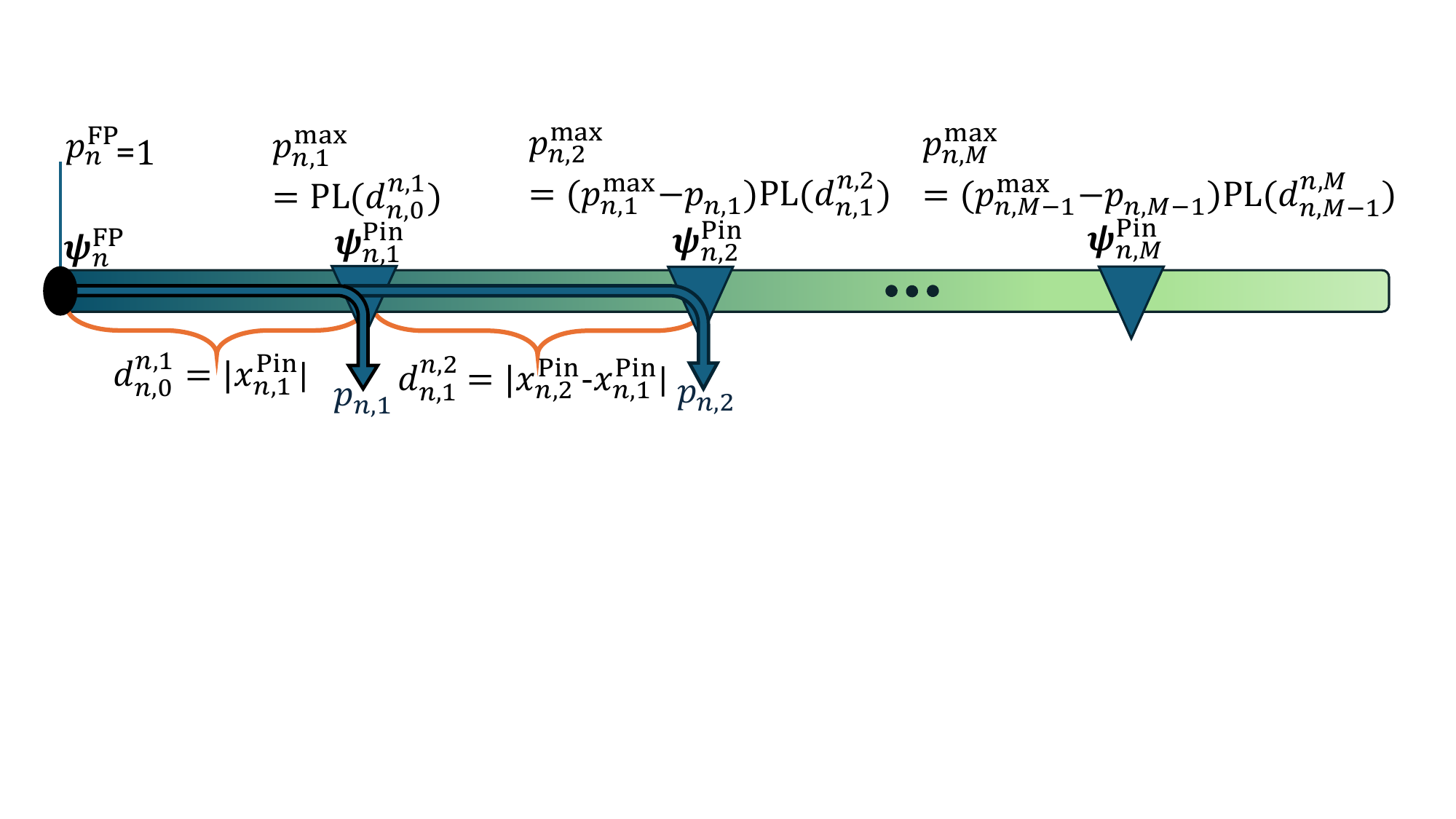}}
\caption{Illustration of the available power ratios among PAs along the $n$-th dielectric waveguide, considering the propagation attenuation effect.}
\label{fig:power}
\end{figure}
Existing studies often oversimplify PA systems, limiting the accuracy and effectiveness of resource allocation strategies. This work proposes a more generalized model~\cite{ding2025flexible, shan2025exploiting, hu2025, xu2025pinching, shan2025multigroup}, which encompasses multiple dielectric waveguides incorporating the beamforming design at the BS, where each PA can independently and dynamically adjust its positions and allocated transmission power ratio, considering waveguide propagation losses. As depicted in Fig.~\ref{fig:power},
the propagation loss between two consecutive PAs, e.g., the $m$-th and ($m-1$)-th PA, in the $n$-th waveguide is denoted as ``PL($d^{n,m}_{n,m-1}$)'' with their spatial separation $d_{n,m-1}^{n,m} = x_{n,m}^{\rm Pin} - x_{n,m-1}^{\rm Pin}$.
The ratio of power allocated to the $m$-th PA in the $n$-th waveguide is denoted by $p_{n,m}$, which can be dynamically managed\footnote{
The ratio of power extracted by each PA depends on its coupling length along with the waveguide~\cite{wang2025modeling, okamoto2021fundamentals}. To abstract from specific hardware implementations, we assume that PAs dynamically adjust their coupling lengths via an auto-scaling mechanism, enabling precise control of it.} as well as its spatial position\footnote{Given the small physical length of PAs relative to practical waveguides~\cite{okamoto2021fundamentals}, their impact on propagation loss is neglected.}, subject to constraints $0 \leq x_{n,m}^{\rm{Pin}} \leq L, d_{n,m-1}^{n,m} \geq \gamma$, where $\gamma > 0$ denotes the minimum allowable spacing among PAs to mitigate coupling effects. Considering lossy dielectric waveguides~\cite{hu2025} and input power $|w_n|^2, \forall n$, at the feeding point of the $n$-th waveguide, the maximum available power coefficient at the $m$-th PA of the $n$-th waveguide, can be mathematically modeled as follows:
\begin{align}\label{eq:power}
p^{\rm{max}}_{n,m} 
=&\; e^{-2 \alpha \left\|\bm{\psi}_n^{\rm FP}-\bm{\psi}_{n,m}^{\rm Pin}\right\|} \nonumber\\
&- \sum_{t=1}^{m-1} p_{n,t} \, 
   e^{-2\alpha \left\|\bm{\psi}_{n,m}^{\rm Pin}-\bm{\psi}_{n,t}^{\rm Pin}\right\|}, \ \forall n,m
\end{align}where $\epsilon_r, \text{tan}\left(\delta_e\right),\alpha = \lambda_g \epsilon_r \pi f_c^2c^{-2} \text{tan}\left(\delta_e\right)\in \mathbb{R}^+$ are dielectric constant, effective electric loss tangent of the dielectric waveguide, and the attenuation constant~\cite{hu2025, balanis2024balanis}, respectively.
Thus, the maximum allowable power ratio for each PA is jointly influenced by its position along the waveguide and the power allocated to preceding PAs. 
Particularly, the introduced model inherently captures critical trade-offs: the balance between waveguide propagation losses and the path losses from PAs to users, and power allocation among the PAs. These comprehensive considerations significantly enhance the practical relevance and effectiveness of subsequent optimization.

\subsection{Channel Model and Performance Metric}
The proposed system assumes a static channel model with perfect channel state information at the BS~\cite{wang2025modeling,hu2025,shan2025exploiting}. The wireless channel between the $M$ PAs on the $n$-th waveguide and the $k$-th user, $\textbf{h}_{n,u_{k}} \hspace{-2.3mm} \in \hspace{-1.4mm} \mathbb{C}^{\hspace{-0.4mm}M \hspace{-0.4mm} \times \hspace{-0.4mm}1}\hspace{-0.9mm}$, is modeled by exploiting the spherical-wave near-field channel model as~\cite{hu2025}:
\begin{align}\label{eq:channel_user}
\mathbf{h}_{n,u_k}(\mathbf{x}^{\rm Pin}_{n})
= \Biggl[ &
\frac{\eta^{1/2}\, e^{-j\frac{2\pi}{\lambda}\|\bm{\phi}_k-\bm{\psi}^{\rm Pin}_{n,1}\|}}
     {\|\bm{\phi}_k-\bm{\psi}^{\rm Pin}_{n,1}\|}, \ldots, \nonumber\\
&\frac{\eta^{1/2}\, e^{-j\frac{2\pi}{\lambda}\|\bm{\phi}_k-\bm{\psi}^{\rm Pin}_{n,M}\|}}
     {\|\bm{\phi}_k-\bm{\psi}^{\rm Pin}_{n,M}\|}
\Biggr]^T, \ \forall n,
\end{align}
where $\textbf{x}^{\rm{Pin}}_{n} \triangleq \{x_{n,1}^{\rm{Pin}}, ..., x_{n,M}^{\rm{Pin}}\}$ denotes the set of PA locations on the $n$-th waveguide and $\eta = \frac{c^2}{16\pi^2f_c^2}$ with $c, f_c,$ and $ \lambda$ representing the light speed, carrier frequency, and wavelength in free space, respectively. 
The  associated phase-shift vector, excluding propagation losses, is defined as~\cite{bereyhi2025downlink, ding2025flexible}:
\begin{equation}\label{eq:channel_prop}
\textbf{g}_n\hspace{-1mm} \left(\hspace{-0.5mm}\textbf{x}^{\rm{Pin}}_{n}\hspace{-0.5mm}\right)\hspace{-1mm}=\hspace{-1mm} \left[\hspace{-0.3mm}e^{-j\frac{2\pi}{\lambda_g}\hspace{-0.5mm}\left\|\hspace{-0.4mm}\bm{\psi}_n^{\rm{FP}} \hspace{-0.8mm}-\hspace{-0.2mm} \bm{\psi}_{n,1}^{\rm{Pin}}\hspace{-0.6mm}\right\|}, ..., e^{-j\frac{2\pi}{\lambda_g}\hspace{-0.5mm}\left\|\hspace{-0.4mm}\bm{\psi}_n^{\rm{FP}} \hspace{-0.8mm}-\hspace{-0.2mm} \bm{\psi}_{n,M}^{\rm{Pin}}\hspace{-0.6mm}\right\|}\hspace{-0.3mm}\right]^{\hspace{-0.5mm}T}\hspace{-1.4mm}, \hspace{0.2mm}\forall n,
\end{equation}
where $\lambda_g = \frac{\lambda}{\eta_{\mathrm{eff}}}$,$\lambda = \frac{c}{f_c}$ and $\eta_{\mathrm{eff}} > 1$ represents the effective refractive index~\cite{ding2025flexible}. 
The propagation losses within the waveguide are separately modeled by leveraging a diagonal power ratio matrix $\mathbf{P}_n \in \mathbb{R}^{M \times M}$, whose diagonal entries $\sqrt{p_{n,m}}, \forall n,m$, representing the power ratio amplitude scaling factor applied to the corresponding PA. The transmission power ratio $p_{n,m}$ is constrained by a maximum allowable power coefficient, denoted by $p^{\max}_{n,m}$, as shown in~\eqref{eq:power}.

To simplify notation, we define the equivalent channel vector from the $N$ waveguides to the $k$-th user, $\forall k$, as:
\begin{align}\label{eq:beta_def}
\bm{\beta}_k
= \Bigl[ &
\mathbf{h}_{1,u_k}^T(\mathbf{x}^{\rm Pin}_1)\,\mathbf{P}_1 \mathbf{g}_1(\mathbf{x}^{\rm Pin}_1), \ldots, \nonumber\\
& \mathbf{h}_{N,u_k}^T(\mathbf{x}^{\rm Pin}_N)\,\mathbf{P}_N \mathbf{g}_N(\mathbf{x}^{\rm Pin}_N)
\Bigr]^T.
\end{align}
Then, the received signal at the $k$-th user can be expressed as:
\begin{equation}\label{rec_signal}
r_k = \bm{\beta}_k^H \textbf{w} s+ \epsilon_k, \ \forall k, 
\end{equation}
where the additive white Gaussian noise (AWGN) at the $k$-th user is denoted as $\epsilon_k \sim \mathcal{CN}\left(0, \sigma_{k}^2\right)$ and $\sigma_{k}^2$ denotes its noise power.
Thus, the achievable data rate for the $k$-th user can be written as:
\begin{equation}\label{eq:rate}
R_k = \log_2\left(1 + \tfrac{\left|\bm{\beta}_k^H \textbf{w}\right|^2}{\sigma_k^2}\right), \quad \forall k.
\end{equation}

\section{Problem Formulation}
The joint optimization problem for maximizing the minimum user data rate is formulated as follows:
\begin{align}
&\underset{\textbf{w}, \ \textbf{x}^{\rm{Pin}}_n, \ \textbf{P}_n} {\rm{maximize}} \hspace{2mm} \underset{k \in \mathcal{K}}{\rm{min}} \{R_k\} \nonumber \\
\rm{s.t.}\ &({\rm C1}): 0 \leq x_{n,m}^{\rm{Pin}} \leq L, \ \forall n, m, \nonumber \\
&({\rm C2}): x_{n,m+1}^{\rm{Pin}} - x_{n,m}^{\rm{Pin}} \geq \gamma, \ \forall n,m, \nonumber \\
&({\rm C3}):0 \leq p_{n,m} \leq p^{\rm{max}}_{n,m}, \ \forall n,m, \nonumber \\
&({\rm C4}): ||\textbf{w}||^2 \leq P_{\rm max}, \label{eq:formulation}
\end{align}
where constraints $({\rm C1})$ and {$({\rm C2})$} ensure that the PA positions and the distance between two adjacent PAs remain within the physical length $L$, and minimum spacing $\gamma$, respectively. Constraints $(\rm C3)$ and  $({\rm C4})$ ensure that the allocated power ratio for PAs and waveguides are within the maximum available values, respectively.
It can be observed from~\eqref{eq:formulation} that the three optimization variables are intricately coupled with each other, which is generally intractable. To address this challenge, we adopt a BCD-based algorithmic framework to obtain an effective low-complexity suboptimal solution. 

\section{Problem Solution}
Motivated by the monotonic properties of the logarithmic function, we introduce an auxiliary optimization variable $t$, representing the minimal signal-to-noise ratio (SNR) among all users. Thus, the original problem in~\eqref{eq:formulation} can be equivalently transformed into its hypograph form as:
\begin{align}\label{eq:reformulation1}
&\underset{t, \,\mathbf{w}, \,\mathbf{x}^{\rm Pin}_n, \,\mathbf{P}_n}{\rm maximize} 
\quad t \nonumber\\
{\rm s.t.}\quad 
&({\rm C1})-({\rm C4}),\; ({\rm C5}):\;
t \leq \left|\bm{\beta}_k^H \mathbf{w}\right|^2 \sigma_k^{-2}, \ \forall k.
\end{align}

Next, we present the three subproblems obtained by applying the BCD method to iteratively and alternately optimize the variable sets $\{t, \textbf{w}\}$, $\{t, \textbf{P}_n\}$, and $\{t, \textbf{x}_n^{\rm Pin}\}$, respectively.

\subsection{Sub-problem $1$: Beamforming Design}
This section addresses the first subproblem of our proposed BCD framework, which can be represented as:
\begin{align}\label{eq:sub_beam}
\underset{t,\, \mathbf{w}}{\rm maximize} \quad t \quad
{\rm s.t.}\quad ({\rm C4}), \ ({\rm C5}).
\end{align}
Noting that this standalone max-min beamforming problem for broadcast systems is inherently NP-hard~\cite{tran2013conic}, we employ a MM approach to derive an iterative algorithm with guaranteed convergence properties\footnote{While semidefinite relaxation can also address this problem, it may not guarantee feasibility in large-scale antenna systems ~\cite{tran2013conic}.}. Specifically, it always converges to a locally optimal solution with polynomial-time computational complexity~\cite{tran2013conic}.
To this end, we first transform the beamforming design problem into an equivalent problem: 
\begin{align}\label{eq:for_beamfor_SOCP}
&\underset{t,\, \mathbf{w},\, \{\hat{r}_k,\, \overline{r}_k,\, \forall k\}}{\rm maximize} 
\quad t \nonumber\\
{\rm s.t.}\quad 
&({\rm C4}),\; \overline{({\rm C5})}: \ \hat{r}_k^2 + \overline{r}_k^2 \geq t\sigma_k^2, \ \forall k, \nonumber\\
&({\rm C6}): \ \hat{r}_k = \Re(\bm{\beta}_k^H \mathbf{w}), \;
\overline{r}_k = \Im(\bm{\beta}_k^H \mathbf{w}), \ \forall k,
\end{align}
where $\hat{r}_k, \overline{r}_k, \forall k$, are auxiliary optimization variables. While $({\rm C6})$ is affine, $\overline{({\rm C5})}$ is recognized as a difference of convex (D.C.) constraint~\cite{yu2021irs}. Therefore, we utilize the MM approach and establish a convex subset of constraint $\overline{({\rm C5})}$:
\begin{equation}
\overline{\overline{({\rm C5})}}: 2\hat{r}_k^{(i_1)} \hat{r}_k \hspace{-0.12mm}+\hspace{-0.12mm} 2\overline{r}_k^{(i_1)} \overline{r}_k \hspace{-0.12mm}-\hspace{-0.12mm} \sigma_k^{\hspace{-0.3mm}2} t \hspace{-0.5mm} \geq \hspace{-0.3mm} \bigr(\hspace{-0.3mm}\hat{r}_k^{(i_1)}\hspace{-0.3mm}\bigr)^{\hspace{-0.55mm}2} \hspace{-0.7mm}+\hspace{-0.23mm} \bigr(\hspace{-0.3mm}\overline{r}_k^{(i_1)}\hspace{-0.3mm}\bigr)^{\hspace{-0.55mm}2} \hspace{-1mm}, \forall k,
\end{equation}
where $\overline{\overline{({\rm C5})}} \Rightarrow \overline{({\rm C5})}$; $\hat{r}_k^{(i_1)}$ and $ \overline{r}_k^{(i_1)}$ denote the respective values obtained at the $i_1$-th iteration of the MM algorithm. Then, a suboptimal solution to~\eqref{eq:for_beamfor_SOCP} is obtained by solving:
\begin{align}
\underset{t, \textbf{w}, \{r_k, a_k, \forall k\}}{\rm{maximize}} \quad t \quad \text{s.t.} \ ({\rm C4}), ({\rm C6}), \overline{\overline{({\rm C5})}}, 
\label{eq:fin_beamfor_SOCP}
\end{align}
which is convex and can be efficiently solved by convex optimization solvers, such as CVX.

\subsection{Sub-problem $2$: PAs' Power Ratio Allocation}
Here, we denote {\small$\textbf{p}_n \hspace{-1.4mm}= \hspace{-0.9mm}\text{Diag}\hspace{-0.7mm}\left(\hspace{-0.2mm}\textbf{P}_{\hspace{-0.35mm}n}\hspace{-0.3mm}\right) \hspace{-0.9mm} \in \hspace{-0.9mm} \mathbb{R}^{\hspace{-0.2mm} M \hspace{-0.4mm}\times \hspace{-0.4mm} 1}\hspace{-0.4mm}, \hspace{-0.3mm}\forall \hspace{-0.1mm} n$}, as the PAs' power ratio vector, and {\small$\bar{\textbf{h}}_{\hspace{-0.3mm}n\hspace{-0.3mm},u_{\hspace{-0.3mm}k}} \hspace{-0.8mm}\left(\hspace{-0.5mm}\textbf{x}_n^{\rm Pin}\hspace{-0.5mm}\right) \hspace{-0.7mm}\in \hspace{-0.7mm}\mathbb{C}^{\hspace{-0.5mm}M \hspace{-0.5mm}\times\hspace{-0.5mm} 1}$} as the equivalent effective channel for $k$-th user from the $n$-th waveguide, where the $m$-th element of which can be represented as {\small$\frac{\eta^{1/2} \hspace{-0.3mm} w_n \hspace{-0.3mm} e^{\hspace{-0.3mm}-j \left(\hspace{-0.5mm} \frac{2\pi}{\lambda} \hspace{-0.3mm}\left\|\bm{\phi}_k \hspace{-0.3mm}- \hspace{-0.3mm}\bm{\psi}^{\rm{Pin}}_{n,m}\hspace{-0.3mm}\right\| + \hspace{-0.3mm}{\frac{2\pi}{\lambda_g}\left\|\hspace{-0.3mm}\bm{\psi}_n^{\rm{FP}} \hspace{-0.3mm}- \hspace{-0.3mm} \bm{\psi}_{n,m}^{\rm{Pin}}\hspace{-0.3mm}\right\|}\right)}}{\left\|\bm{\phi}_k - \bm{\psi}^{\rm{Pin}}_{n,m}\right\|}, \forall m$}.
Therefore, the power ratio optimization problem can be formulated as:
\begin{align}\label{eq:sub_formulation2}
\underset{t, \ \{\textbf{p}_n, \forall n\}}{\rm{maximize}} &\quad t \nonumber\\
{\rm s.t.}\quad 
({\rm C3}), ({\rm C5}):\ &\ t\sigma_k^2 \leq 
\Bigl|\sum_{n=1}^N \bar{\textbf{h}}^H_{n,u_k} \textbf{p}_n\Bigr|^2, \ \forall k.
\end{align}

Noting that the problem in~\eqref{eq:sub_formulation2} remains non-convex due to the quadratic terms in $({\rm C5})$, we employ the MM to establish a global lower bound for the right-hand side of $({\rm C5})$ as:

\begin{align}
\Bigl|\sum_{n\in\mathcal{N}}\bar{\mathbf{h}}^H_{n,u_k}\,\mathbf{p}_n\hspace{-0.5mm}\Bigr|^2
      &\hspace{-1mm}\ge\hspace{-1mm} -2\,\Re\Bigl[\bigl(\hspace{-1mm}\sum_{n\in\mathcal{N}}\bar{\mathbf{h}}^H_{n,u_k}\,\mathbf{p}_n^{(i_2)}\bigr)^*      \sum_{n\in\mathcal{N}}\bar{\mathbf{h}}^H_{n,u_k}\,\mathbf{p}_n\Bigr] \nonumber \\
      &\quad +\Bigl|\sum_{n\in\mathcal{N}}\bar{\mathbf{h}}^H_{n,u_k}\,\mathbf{p}_n^{(i_2)}\Bigr|^2,
      \quad\forall k,
\end{align}
where $\textbf{p}_n^{(i_2)}$ denote the respective value obtained at the $i_2$-th iteration of the MM algorithm.
As such, a convex subset of $({\rm C5})$, with $\widetilde{({\rm C5})} \Rightarrow ({\rm C5}), \forall k$, can be derived as:
\begin{align}\label{eq:C5_tilde}
\widetilde{({\rm C5})}: \quad 
&+ 2\Re \Biggl[ 
   \Biggl(\sum_{n \in \mathcal{N}} \bar{\textbf{h}}^H_{n,u_k}\, 
   \textbf{p}_n^{(i_2)} \Biggr)^{*} 
   \sum_{n \in \mathcal{N}} \bar{\textbf{h}}^H_{n,u_k}\,\textbf{p}_n
   \Biggr] \nonumber\\
& -\Biggl|\sum_{n \in \mathcal{N}} \bar{\textbf{h}}^H_{n,u_k} \, 
   \textbf{p}_n^{(i_2)} \Biggr|^2  -\hspace{-0.5mm} t\sigma_k^2  \;\geq\; 0.
\end{align}
Thus, a suboptimal solution to~\eqref{eq:sub_formulation2} is obtained by solving:
\begin{eqnarray}\label{eq: power_fin}
\underset{t, \ \{\textbf{p}_n, \forall n \} }{\rm{maximize}} \hspace{2mm}t \quad {\rm{s.t.}}\ ({\rm C3}), \widetilde{({\rm C5})},
\end{eqnarray}
which is convex and can be solved with CVX.


\subsection{Sub-problem $3$: PAs' Location}
With fixed $\textbf{w}$ and $\textbf{p}_n, \forall n,$ the PAs' location optimization problem can be formulated as:
\begin{equation}\label{eq:sub_formulation3}
\begin{aligned}
&\underset{t, \ \textbf{x}_n^{\rm Pin}}{\rm{maximize}} \hspace{2mm} t \ \ \rm{s.t.}\ ({\rm C1}), ({\rm C2}),\\
&(\hspace{-0.1mm}{\rm C5})\hspace{-0.8mm}:\hspace{-0.8mm} \sigma_k^2 t \hspace{-0.8mm}\leq\hspace{-1mm} \left|\hspace{-0.4mm}\sum_{n,m =1}^{N,M} \hspace{-1.5mm} \left[\hspace{-0.8mm} \tfrac{w_n\hspace{-0.3mm}\eta^{\frac{1}{2}} \hspace{-0.6mm} p_{n,m}^{\frac{1}{2}} \hspace{-0.3mm} e^{\hspace{-0.6mm}-\hspace{-0.5mm}j \hspace{-0.5mm}\left(\hspace{-0.9mm} \frac{2\pi} {\lambda} \hspace{-0.5mm} \left\|\hspace{-0.5mm} \bm{\phi}_k \hspace{-0.5mm}- \hspace{-0.5mm}\bm{\psi}^{\rm{Pin}}_{n,m}\hspace{-0.5mm}\right\| \hspace{-0.3mm}+\hspace{-0.3mm} {\frac{2\pi}{\lambda_g} \hspace{-0.5mm} \left\| \hspace{-0.4mm} \bm{\psi}_n^{\rm{FP}} \hspace{-0.9mm}- \hspace{-0.2mm}\bm{\psi}_{n,m}^{\rm{Pin}}\hspace{-0.6mm}\right\|}\hspace{-0.5mm}\right)} }{\left\|\bm{\phi}_k - \bm{\psi}^{\rm{Pin}}_{n,m}\right\|} \hspace{-0.7mm}\right] \hspace{-0.5mm} \right|^2\hspace{-1.8mm}, \hspace{-0.6mm}\forall \hspace{-0.2mm}k.
\end{aligned}
\end{equation}
It can be noted that the PA locations simultaneously influence both phase alignment and path loss. Indeed, this inherent coupling makes the optimization a challenging amplitude-dependent phase problem.
To facilitate tractability, we define $\textbf{C}_{n} \hspace{-1.2mm}= \text{diag}\hspace{-1mm}\left(\hspace{-0.8mm}w_n \eta^{\frac{1}{2}} p_{n,m}^{\frac{1}{2}}, ..., w_n \eta^{\frac{1}{2}} p_{n,M}^{\frac{1}{2}}\hspace{-1mm}\right) \in \mathbb{C}^{M \times M}$, $\textbf{f}_{n,u_k} \hspace{-1.2mm}=\hspace{-1.2mm} \left[\tfrac{1}{\left\|\bm{\phi}_k - \bm{\psi}^{\rm{Pin}}_{n,1}\right\|}, ..., \tfrac{1}{\left\|\bm{\phi}_k - \bm{\psi}^{\rm{Pin}}_{n,M}\right\|} \right]^T \in \mathbb{R}^{M \times 1} $, and $\textbf{a}^H_{n, u_k} \hspace{-1.7mm} =\hspace{-1.5mm}\left[ e^{-j \theta_{n,1,k}}, ...,  e^{-j \theta_{n,M,k}} \hspace{-0.7mm}\right] \in \mathbb{C}^{1 \times M}, \forall n, k$, where $\theta_{n,m,k} \hspace{-1mm} = \hspace{-1mm} \frac{2\pi}{\lambda}\left\| \bm{\phi}_k \hspace{-0.7mm}-\hspace{-0.7mm} \bm{\psi}^{\rm{Pin}}_{n,m}\right\| + {\frac{2\pi}{\lambda_g}\left\|\bm{\psi}_n^{\rm{FP}} \hspace{-1mm}-\hspace{-0.7mm} \bm{\psi}_{n,m}^{\rm{Pin}}\right\|}, \forall n,m,k$. Moreover, we let $\textbf{a}^H_{k} \hspace{-1mm}=\hspace{-1mm} \left[\textbf{a}^H_{1, u_k}, ...,  \textbf{a}^H_{N, u_k}\right] \in \mathbb{C}^{1 \times MN}$, $\textbf{C} \hspace{-0.7mm}=\hspace{-0.7mm} \text{blkdiag}\left(\textbf{C}_{1}, ..., \textbf{C}_{N}\right) \in \mathbb{C}^{MN \times MN}$, and $\textbf{f}_k = \text{vec}\left([\textbf{f}_{1,u_k}, ..., \textbf{f}_{N,u_k}]\right) \in \mathbb{R}^{MN \times 1}, \forall k$. 
With defining $\textbf{A}_k = \textbf{a}_k \textbf{a}_k^H$ and $\textbf{F}_k = \textbf{f}_k\textbf{f}_k^H, \forall k$, $({\rm C5})$ can be equivalently expressed by exploiting the Frobenius norm, as:
\begin{align}
({\rm C5}) \hspace{-1mm}:\hspace{-1mm}  \left\| \textbf{F}_{\hspace{-0.5mm}k} \hspace{-0.8mm}+\hspace{-0.8mm} \textbf{C}^{\hspace{-0.2mm}H} \hspace{-0.99mm}\textbf{A}_k \hspace{-0.5mm} \textbf{C}  \right\|_F^2 
\hspace{-1.6mm}-\hspace{-0.7mm} \left\| \textbf{F}_{\hspace{-0.5mm}k} \right\|_F^2 
\hspace{-1mm}-\hspace{-0.7mm} \left\| \textbf{C}^{\hspace{-0.2mm}H} \hspace{-1.1mm} \textbf{A}_k \hspace{-0.3mm} \textbf{C} \right\|_F^2
\geq 2\sigma_k^2 t, \forall k,
\end{align}
which reveals a D.C. format. Thus, the problem can be equivalently reformulated by using the auxiliary variables, $\textbf{F}_k$ and $\textbf{A}_k$, to characterize the pure path loss and phase alignment components, respectively, as follows:
\begin{align}\label{eq:sub_formulation3_1}
&\underset{t, \hspace{0.5mm} x_{n,m}^{\rm Pin}, \hspace{0.5mm} \textbf{F}_k, \hspace{0.5mm} \textbf{A}_k,\hspace{0.5mm} \theta_{l(n,m),k}}{\rm{maximize}} \hspace{2mm} t \ \ \rm{s.t.}\ ({\rm C1}), ({\rm C2}), ({\rm C5}), \nonumber \\
&({\rm C7}): \left(x_{n,m}^{\rm Pin} - x_k\right)^2= \frac{1}{\text{Diag} (\textbf{F}_k)_{l(n,m)}} - \hat{s}_{n,k}, \ \forall n, m, k, \nonumber \\
&({\rm C8}): [\hspace{-0.3mm}\textbf{A}_k\hspace{-0.3mm}]_{\hspace{-0.3mm}l\hspace{-0.4mm}(\hspace{-0.4mm}n,m\hspace{-0.4mm}), l(\hspace{-0.4mm}n'\hspace{-0.4mm},m'\hspace{-0.4mm})} \hspace{-1mm}=\hspace{-1mm} e^{\hspace{-1mm}-\hspace{-0.3mm}j\hspace{-0.3mm}\left(\hspace{-0.6mm} \theta_{\hspace{-0.3mm}l\hspace{-0.3mm}(\hspace{-0.3mm}n,m\hspace{-0.3mm})\hspace{-0.3mm},\hspace{-0.3mm}k} -\theta_{\hspace{-0.3mm}l\hspace{-0.3mm}(\hspace{-0.3mm}n'\hspace{-0.3mm},m'\hspace{-0.3mm}),\hspace{-0.1mm}k\hspace{-0.3mm}} \hspace{-0.3mm} \right)}\hspace{-0.8mm}, \forall \hspace{-0.15mm}n, \hspace{-0.15mm}m\hspace{-0.25mm}, \hspace{-0.15mm}n'\hspace{-0.95mm}, \hspace{-0.15mm}m'\hspace{-0.95mm}, \hspace{-0.15mm}k, \nonumber \\
&({\rm C9}): \theta_{l(n,m),k} \hspace{-0.8mm}=\hspace{-0.8mm} \frac{2\pi}{\lambda} \hspace{-0.99mm} \left(\hspace{-0.5mm}\text{Diag} (\textbf{F}_k)_{l(n,m)}\hspace{-0.6mm}\right)^{\hspace{-0.8mm}-\frac{1}{2}} \hspace{-0.99mm}+ \hspace{-0.8mm}\frac{2 \pi}{\lambda_g}\hspace{-0.4mm} x_{n,m}^{\rm Pin}, \hspace{-0.6mm}\forall \hspace{-0.4mm}n, \hspace{-0.4mm}m, \hspace{-0.4mm}k, \nonumber \\
&({\rm C10})\hspace{-0.4mm}:\hspace{-0.4mm} \textbf{F}_{\hspace{-0.4mm}k}\hspace{-0.4mm},\hspace{-0.4mm} \textbf{A}_{\hspace{-0.4mm}k} \hspace{-0.7mm}\bm{\succeq} \hspace{-0.7mm}\textbf{0}, \hspace{-0.4mm} \forall \hspace{-0.2mm}k,  (\hspace{-0.4mm}{\rm C11}\hspace{-0.4mm})\hspace{-0.9mm}:\hspace{-0.4mm} \text{Rank}\hspace{-0.6mm}\left(\hspace{-0.4mm}\textbf{A}_{\hspace{-0.2mm}k}\hspace{-0.4mm}\right)\hspace{-0.95mm}, \text{Rank}\hspace{-0.6mm}\left(\hspace{-0.3mm}\textbf{F}_{\hspace{-0.4mm}k}\hspace{-0.4mm}\right)\hspace{-0.7mm} =\hspace{-0.7mm} 1\hspace{-0.4mm},\hspace{-0.4mm} \forall \hspace{-0.2mm} k, 
\end{align}
where $l(\hspace{-0.5mm}n,\hspace{-0.5mm}m\hspace{-0.5mm})\hspace{-1mm}=\hspace{-1mm}(\hspace{-0.5mm}n\hspace{-1mm}-\hspace{-1mm}1\hspace{-0.5mm})M\hspace{-0.6mm}+\hspace{-0.3mm}m$ maps indices from two-dimensional to a single-dimensional stacked vector form with $\hat{s}_{n,k}\hspace{-1mm}=\hspace{-1mm}(y_k \hspace{-1mm}-\hspace{-1mm} D_n)^2 \hspace{-1mm}+\hspace{-1mm} d^2$ and $\text{Diag}(\textbf{F}_k)_{l(n,m)}\hspace{-1mm}=\hspace{-1mm}[\textbf{F}_k]_{l(n,m),l(n,m)}$ for simplifying the notation. Specifically, constraints $({\rm C7})$–$({\rm C9})$ establish explicit relationships between the primary variable $x_{n,m}^{\rm Pin}$ and auxiliary variables $\textbf{F}_k$, $\textbf{A}_k$, and $\theta_{n,m,k}$. 

We first address the non-convex equality constraints $({\rm C7})$ and $({\rm C9})$. Taking $({\rm C7})$ as an example, it can equivalently be represented by two inequality constraints:
\begin{align}
({\rm C7a}):& \hspace{-0.3mm}\left(\hspace{-0.6mm}x_{n,m}^{\rm Pin}\hspace{-0.7mm} -\hspace{-0.7mm} x_k\hspace{-0.6mm}\right)^{\hspace{-0.6mm}2} \hspace{-1.4mm}-\hspace{-0.8mm} \tfrac{1}{\text{Diag} (\textbf{F}_k)_{l(n,m)}} \hspace{-0.7mm}+\hspace{-0.5mm} \hat{s}_{n,k} \hspace{-0.8mm} \leq \hspace{-0.8mm} 0 , \forall n,m,k, \nonumber \\  
({\rm C7b}):& \tfrac{1}{\text{Diag} (\textbf{F}_k)_{l(n,m)}} \hspace{-0.8mm}-\hspace{-0.8mm} \hat{s}_{n,k} \hspace{-0.9mm}-\hspace{-0.99mm} \left(\hspace{-0.6mm}x_{n,m}^{\rm Pin}\hspace{-0.7mm} -\hspace{-0.7mm} x_k\hspace{-0.6mm}\right)^{\hspace{-0.6mm}2} \hspace{-0.8mm} \leq \hspace{-0.8mm} 0, \forall n,m,k,
\end{align}
both of which remain non-convex but admit a D.C. structure. Similarly, $({\rm C9})$ is equivalent to:
\begin{align}
(\hspace{-0.5mm}{\rm C9a}\hspace{-0.5mm})\hspace{-0.9mm}:&  \tfrac{2\pi}{\lambda} \hspace{-1.3mm} \left(\hspace{-0.3mm}\text{Diag} \hspace{-0.3mm} (\hspace{-0.3mm}\textbf{F}_{\hspace{-0.3mm}k}\hspace{-0.3mm})_{\hspace{-0.3mm}l\hspace{-0.3mm}(\hspace{-0.3mm}n,m\hspace{-0.3mm})}\hspace{-0.5mm}\right)^{\hspace{-1mm}-\hspace{-0.3mm}\frac{1}{2}} \hspace{-0.9mm}+ \hspace{-0.7mm} \tfrac{2 \pi}{\lambda_g} x_{n,m}^{\rm Pin} \hspace{-0.9mm} - \hspace{-0.7mm} \theta_{l(\hspace{-0.3mm}n,m\hspace{-0.3mm}),k} \hspace{-0.4mm}\leq \hspace{-0.5mm} 0, \hspace{-0.5mm} \nonumber \\
(\hspace{-0.5mm}{\rm C9b}\hspace{-0.5mm})\hspace{-0.9mm}:&  \theta_{l(\hspace{-0.3mm}n,m\hspace{-0.3mm}),k} \hspace{-0.9mm} - \hspace{-0.7mm}  \tfrac{2\pi}{\lambda} \hspace{-1.3mm} \left(\hspace{-0.3mm}\text{Diag} \hspace{-0.3mm} (\hspace{-0.3mm}\textbf{F}_{\hspace{-0.3mm}k}\hspace{-0.3mm})_{\hspace{-0.3mm}l\hspace{-0.3mm}(\hspace{-0.3mm}n,m\hspace{-0.3mm})}\hspace{-0.5mm}\right)^{\hspace{-1mm}-\hspace{-0.3mm}\frac{1}{2}} \hspace{-0.9mm}- \hspace{-0.9mm} \tfrac{2 \pi}{\lambda_g} x_{n,m}^{\rm Pin}  \leq 0, \hspace{-0.5mm}
\end{align}
$\forall \hspace{-0.2mm}n, \hspace{-0.5mm}m, \hspace{-0.5mm}k,$ where constraint $({\rm C9a})$ is convex, whereas constraint $({\rm C9b})$ is a D.C. format. Moreover, to effectively handle the rank-one constraint $({\rm C11})$, we introduce the following lemma:
\begin{lemma}
The rank-one constraints $({\rm C11})$ are equivalent to constraint $(\overline{{\rm C11}})$, given by:
\end{lemma}
\vspace{-1em}
\begin{equation}
(\overline{{\rm C11}}): \ \left\|\textbf{A}_k\right\|_* - \left\|\textbf{A}_k\right\|_2 \leq 0, \ \ \left\|\textbf{F}_k\right\|_* - \left\|\textbf{F}_k\right\|_2 \leq 0, \forall k.
\end{equation}
\begin{proof}
For any $\textbf{X} \in \mathbb{H}^n \succeq \textbf{0}$ , the inequality $\left\|\textbf{X}\right\|_* = \sum_i\Lambda_i \geq \left\|\textbf{X}\right\|_2 = \text{max}\{\Lambda_i\}$ holds, where $\Lambda_i$ is the $i$-th singular value of $\textbf{X}$. Equality holds if and only if $\textbf{X}$ is rank-one.
\end{proof}

Utilizing Lemma 1, constraint $(\overline{{\rm C11}})$ is now represented as a D.C. form.
Given that $\text{Rank}(\textbf{A}_k) = 1, \textbf{A}_k \succeq \bm{0}, \forall k$, the matrix $\textbf{A}_k$ is fully characterized by any non-zero column $1 \leq j \leq MN$. Without loss of generality, we utilize the first column to simplify notation; however, any other non-zero column could serve this purpose. By leveraging trigonometric identities, $({\rm C8})$ can be equivalently transformed into:
\begin{align}
({\rm C8a})\hspace{-0.8mm}:& \Re\left([\textbf{A}_k]_{i,1}\right) = \cos\left(\hat{\theta}_{i,k}\right), \forall i \in \{2, ..., MN\}, \nonumber \\
({\rm C8b})\hspace{-0.8mm}:& \Im\left([\textbf{A}_k]_{i,1}\right) = -\sin\left(\hat{\theta}_{i,k}\right), \forall i \in \{2, ..., MN\},
\end{align}
where $\hat{\theta}_{i,k} = \theta_{i,k} - \theta_{1,k}$ and the index $i$ maps to the pair $(n,m)$ through the relationship $i = (n-1)M + m$.
To further address the intrinsic non-convexity, we introduce a penalized form of the optimization problem \eqref{eq:sub_formulation3_1} as follows~\cite{mao2023amplitude, yu2021irs}:
\begin{align}\label{eq:sub_formulation3_withoutSCA}
  &\hspace{-1.85em}
      \underset{%
        \substack{%
          \hat{\theta}_{i,k}, \theta_{l(n,m),k},\hspace{0.3mm} t,\\
          \textbf{F}_{k}, \ \textbf{A}_{k}, \  x_{n,m}^{\rm Pin}
        }%
      }{\mathrm{maximize}} t
      -\rho 
      \sum_{k=1}^{K}\sum_{i=2}^{MN}
        \Phi\left([\textbf{A}_k]_{i,1}, \hat{\theta}_{i,k}\right)
  \nonumber\\
  \mathrm{s.t.}\;
     &(\hspace{-0.2mm}\mathrm C1\hspace{-0.2mm})\hspace{-0.25mm},\hspace{-0.2mm}(\hspace{-0.2mm}\mathrm C2\hspace{-0.2mm})\hspace{-0.25mm},\hspace{-0.2mm}(\hspace{-0.2mm}\mathrm C5\hspace{-0.2mm})\hspace{-0.25mm},\hspace{-0.2mm}({\rm\hspace{-0.2mm}\mathrm C7a}\hspace{-0.2mm})\hspace{-0.25mm},\hspace{-0.2mm}(\hspace{-0.2mm}\mathrm {\rm C7b}\hspace{-0.2mm})\hspace{-0.25mm},
    \hspace{-0.2mm}(\hspace{-0.2mm}\mathrm {\rm C9a}\hspace{-0.2mm})\hspace{-0.25mm},\hspace{-0.2mm}(\hspace{-0.2mm}\mathrm {\rm C9b}\hspace{-0.2mm})\hspace{-0.25mm},\hspace{-0.2mm}(\hspace{-0.2mm}\mathrm C10\hspace{-0.2mm})\hspace{-0.2mm},\hspace{-0.2mm}\overline{(\hspace{-0.2mm}\mathrm C11\hspace{-0.2mm})}\hspace{-0.2mm},\hspace{-0.2mm}
  \nonumber\\
  &(\hspace{-0.2mm}\mathrm C12\hspace{-0.2mm})\hspace{-0.27mm}\colon
\hspace{-0.27mm}\hat{\theta}_{i,k} \hspace{-0.3mm} =\hspace{-0.3mm} \theta_{i,k} \hspace{-0.2mm}-\hspace{-0.2mm} \theta_{1,k}, \forall k \hspace{-0.4mm}\in \hspace{-0.7mm}\mathcal K, i\hspace{-0.4mm}\in \hspace{-0.7mm}\{2,\dots,MN\},
\end{align}
where {\footnotesize$\Phi\hspace{-0.8mm}\left(\hspace{-0.6mm}[\textbf{A}_k]_{\hspace{-0.3mm}i,\hspace{-0.3mm}1}\hspace{-0.3mm}, \hat{\theta}_{\hspace{-0.3mm}i\hspace{-0.3mm},k}\hspace{-0.85mm}\right) \hspace{-0.85mm}=\hspace{-0.6mm} \left|\hspace{-0.3mm}\Re\hspace{-0.6mm}\left(\hspace{-0.2mm}[\textbf{A}_k]_{i,1}\hspace{-0.6mm}\right) \hspace{-0.6mm}-\hspace{-0.6mm} \cos\hspace{-0.6mm}\left(\hspace{-0.6mm}\hat{\theta}_{i,k}\hspace{-0.6mm}\right) \hspace{-0.6mm}\right|^2 \hspace{-1mm}+\hspace{-0.8mm} \left|\hspace{-0.3mm}\Im\hspace{-0.6mm}\left(\hspace{-0.2mm}[\hspace{-0.2mm}\textbf{A}_k]_{i,1}\hspace{-0.4mm}\right)\hspace{-0.6mm} + \hspace{-0.6mm}\sin\hspace{-0.6mm}\left(\hspace{-0.6mm}\hat{\theta}_{i,k}\hspace{-0.6mm}\right)\hspace{-0.5mm} \right|^2$}, $\forall i, k,$ and $\rho > 0$ is a penalty coefficient that imposes a cost on the violation of equality constraints $({\rm C8a})$ and $({\rm C8b})$, and constraint $({\rm C12})$ establish the relationship between $\theta_{l(n,m),k}$ and $\hat{\theta}_{i,k}$. Specifically, problems ~\eqref{eq:sub_formulation3_1} and in~\eqref{eq:sub_formulation3_withoutSCA} are equivalent when $\rho$ is sufficient large, i.e., $\rho \rightarrow \infty$~\cite{yu2021irs}.

Nonetheless, the optimization problem in \eqref{eq:sub_formulation3_withoutSCA} remains non-convex as constraints $({\rm C5})$, $({\rm C7a})$, $({\rm C7b})$, $({\rm C9b})$, and $\overline{({\rm C11})}$ exhibit a D.C. structure, whereas the penalty term in the objective poses the periodic nature. To address this, we employ the MM algorithm. Given the D.C. structure, we derive global underestimators for the aforementioned non-convex terms using their first-order Taylor series expansions:
    \begin{align}
      \|\hspace{-0.2mm}\mathbf F_{\hspace{-0.2mm}k} \hspace{-0.4mm}+\hspace{-0.2mm} \mathbf C^H \hspace{-0.5mm} \mathbf A_k \hspace{-0.2mm} \mathbf C\|_F^2
       \ge &  \overline{\|\hspace{-0.1mm}\mathbf F_{\hspace{-0.2mm}k} \hspace{-0.5mm}+\hspace{-0.2mm} \mathbf C^H \hspace{-0.5mm} \mathbf A_k \hspace{-0.2mm} \mathbf C\|_F^2} \nonumber\\
       = &\|\mathcal S_{(i_3)}\|_F^2
        \hspace{-0.4mm} + \hspace{-0.4mm} 2\mathrm{Tr}\bigl(\hspace{-0.2mm} \mathcal S_{(i_3)}^H \hspace{-0.1mm} \triangle\mathbf F_k \hspace{-0.1mm}\bigr)
        \hspace{-0.4mm} \nonumber \\ &+\hspace{-0.4mm} 2 \mathrm{Tr}\hspace{-0.0mm}\bigl(\hspace{-0.2mm}\mathcal S_{\hspace{-0.0mm}(\hspace{-0.0mm}i_3\hspace{-0.0mm})}^{\hspace{-0.2mm}H} \hspace{-0.4mm} \mathbf C^{\hspace{-0.2mm}H} \hspace{-0.5mm} \triangle\mathbf A_k \hspace{-0.2mm}\mathbf C\bigr)\hspace{-0.1mm},\hspace{-0.0mm}
     \forall k,
\end{align}
\begin{align}
    \left( \hspace{-0.0mm}x_{ \hspace{-0.0mm}n\hspace{-0.0mm},\hspace{-0.0mm}m}^{\text{Pin}}\hspace{-0.2mm} - \hspace{-0.0mm} x_{\hspace{-0.0mm}k} \hspace{-0.0mm}\right)^{\hspace{-0.0mm}2} \hspace{-0.2mm} \ge \hspace{-0.2mm} x_{\hspace{-0.1mm}n\hspace{-0.0mm},\hspace{-0.0mm}m\hspace{-0.0mm},\hspace{-0.0mm}k}^{\rm aff} \hspace{-0.2mm} = & 2 \hspace{-0.2mm}\left( \hspace{-0.5mm}x_{n,m}^{\hspace{-0.0mm}\text{Pin}(\hspace{-0.1mm}i_3\hspace{-0.1mm})} \hspace{-0.5mm} - \hspace{-0.5mm} x_k \hspace{-0.5mm}\right)  \hspace{-0.7mm}\left( \hspace{-0.5mm}x_{n,m}^{\text{Pin}} \hspace{-0.5mm} - \hspace{-0.5mm} x_{n,m}^{\text{Pin}(i_3)} \hspace{-0.5mm}\right)\hspace{-0.5mm}, \nonumber \\
    & + \left( \hspace{-0.5mm}x_{n,m}^{\hspace{-0.0mm}\text{Pin}(\hspace{-0.1mm}i_3\hspace{-0.1mm})} \hspace{-0.7mm} - \hspace{-0.5mm} x_k \hspace{-0.5mm}\right)^{ \hspace{-0.5mm}2} \forall \hspace{-0.1mm}n,\hspace{-0.3mm}m,\hspace{-0.3mm}k,
\end{align}
\begin{align}
    \frac{1}{\text{Diag}(\textbf{F}_{\hspace{-0.0mm}k}\hspace{-0.0mm})_{\hspace{-0.0mm}l\hspace{-0.0mm}(\hspace{-0.0mm}n,m)}} \hspace{-0.5mm} \ge \hspace{-0.5mm} z_{n,m,k}^{\rm aff} \hspace{-0.5mm} = &- \hspace{-0.2mm} \tfrac{1}{\left(z_{n,m,k}^{(i_3)}\right)^2} \hspace{-0.5mm} \left(\hspace{-0.5mm}z_{n,m,k} \hspace{-0.2mm} - \hspace{-0.2mm} z_{n,m,k}^{(i_3)}\hspace{-0.4mm}\right)\hspace{-0.5mm} \nonumber \\
    &+\hspace{-0.5mm} \tfrac{1}{z_{n,m,k}^{(i_3)}} \hspace{-0.2mm}, \hspace{-0.1mm}\forall \hspace{-0.0mm}n,\hspace{-0.0mm}m,\hspace{-0.0mm}k,
\end{align}
\begin{align}
    \frac{2\pi}{\lambda} z_{n,m,k}^{-1/2} \ge \overline{z_{n,m,k}^{\rm aff}} = & - \frac{\pi}{\lambda} \left(z_{n,m,k}^{(i_3)}\right)^{-\frac{3}{2}}\left(z - z_{n,m,k}^{(i_3)}\right), \nonumber \\
    &\frac{2\pi} {\lambda}\left(z_{n,m,k}^{(i_3)}\right)^{-\frac{1}{2}} \forall n,m,k,
\end{align}
\begin{align}
\left\| \textbf{X}_{ k} \right\|_{ 2} \hspace{-0.3mm} \geq \hspace{-0.3mm}\left\| \textbf{X}^{\rm aff}_k\right\|_{2}\hspace{-0.3mm} =&\text{Tr}\hspace{-0.3mm}\left[\bm{\Phi}_{\rm max}\hspace{-1mm}\left(\hspace{-0.7mm}\textbf{X}_k^{(i_3)}\hspace{-0.7mm}\right) \hspace{-1mm}\bm{\Phi}^H_{\rm max}\hspace{-1mm}\left(\hspace{-0.7mm}\textbf{X}_k^{(i_3)}\hspace{-0.7mm}\right) \hspace{-1mm}\left(\hspace{-0.5mm}\textbf{X}_k \hspace{-0.75mm}-\hspace{-0.7mm} \textbf{X}_k^{(i_3)}\hspace{-0.7mm}\right)\hspace{-0.5mm} \right] \hspace{-0.88mm} \nonumber \\
&+\hspace{-0.2mm}  \left\|\textbf{X}_k^{(i_3)}\hspace{-0.1mm}\right\|_{2}\hspace{-0.2mm}, \forall k, \textbf{X},
\end{align}
where $\textbf{X} \in \{\textbf{A}, \textbf{F}\}$. To simplify the notation, we denote $\text{Diag}(\textbf{F}_k)_{l(n,m)} = z_{n,m,k}$ and  $\mathcal{S}_{(i_3)} = \textbf{F}_k^{(i_3)} + \textbf{C}^H \textbf{A}_k^{(i_3)} \textbf{C} $, $\bm{\triangle} \textbf{F}_k = \textbf{F}_k - \textbf{F}_k^{(i_3)}$, and $\bm{\triangle} \textbf{A}_k = \textbf{A}_k - \textbf{A}_k^{(i_3)}$. The terms with ``$(i_3)$'' are the solutions obtained in the $i_3$-th iteration for the respective variables. 
Similarly, we apply the MM technique along with a Lipschitz gradient surrogate to establish a valid global upper bound~\cite{mairal2015incremental} for $\Phi\left([\textbf{A}_k]_{i,1}, \hat{\theta}_{i,k}\right)$: 
\begin{align}
      \overline{\Phi}^{(i_3)}_{i,k} 
      = &  2\left(\hat{\Re}_{i,k}^{(i_3)}\sin \hat{\theta}_{i,k}^{(i_3)} \hspace{-0.2mm}+\hspace{-0.2mm} \hat{\Im}_{i,k}^{(\hspace{-0.3mm}i_3\hspace{-0.3mm})}\hspace{-0.7mm}\cos \hat{\theta}_{i,k}^{(\hspace{-0.3mm}i_3\hspace{-0.3mm})}\hspace{-0.8mm}\right)\hspace{-0.8mm}\triangle \hat{\theta}_{i,k} \nonumber \\
      &+ \frac{L_\text{AR}}{2}\triangle \Re_{i,k} + \frac{L_\text{AI}}{2}\triangle \Im_{i,k} + \frac{L_{\rm TH}}{2} \left(\triangle \hat{\theta}_{i,k}\right)^2 \nonumber \\
      & + 2\hat{\Re}_{i,k}^{(i_3)}\hspace{-0.1mm}\triangle \Re_{i,k}\hspace{-0.2mm}+\hspace{-0.3mm} 2\hat{\Im}_{i,k}^{(i_3)}\triangle \Im_{i,k} + \Phi^{(i_3)} , \ \forall i, k,
\end{align}
where {\small $\hat{\Re}_{i,k}^{(\hspace{-0.3mm}i_3\hspace{-0.3mm})} \hspace{-0.5mm}= \hspace{-0.5mm}\Re([\textbf{A}_k]_{i,1}^{(\hspace{-0.3mm}i_3\hspace{-0.3mm})}) \hspace{-0.5mm}-\hspace{-0.5mm} \cos\hspace{-0.5mm} \hat{\theta}_{i,k}^{(\hspace{-0.3mm}i_3\hspace{-0.3mm})}$, $\hat{\Im}_{i,k}^{(\hspace{-0.3mm}i_3\hspace{-0.3mm})} \hspace{-0.5mm}=\hspace{-0.5mm} \Im([\textbf{A}_k]_{i,1}^{(\hspace{-0.3mm}i_3\hspace{-0.3mm})}) \hspace{-0.5mm}+\hspace{-0.5mm} \sin\hspace{-0.5mm} \hat{\theta}_{i,k}^{(\hspace{-0.3mm}i_3\hspace{-0.3mm})}$, $\triangle \hat{\theta}_{i,k} \hspace{-0.5mm}=\hspace{-0.5mm} \hat{\theta}_{i,k} \hspace{-0.5mm}- \hat{\theta}_{i,k}^{(\hspace{-0.3mm}i_3\hspace{-0.3mm})}$, $\triangle \Re_{i,k}=\Re([\textbf{A}_k]_{i,1}) - \Re([\textbf{A}_k]_{i,1}^{(i_3)}) $, $\triangle \Im_{i,k} = \Im([\textbf{A}_k]_{i,1}) - \Im([\textbf{A}_k]_{i,1}^{(i_3)})$}, with the Lipschitz constants $L_{\rm AR} = L_{\rm AI} = 2$ and $L_{\rm TH} = 4$ from the maximum curvature~\cite{mairal2015incremental}.

Therefore, the convex subsets of the above non-convex constraint sets are given by:
\begin{align}
\widetilde{\overline{({\rm C5})}}: &\tfrac{\overline{\left\| \textbf{F}_k +  \textbf{C}^{\hspace{-0.3mm}H}\hspace{-0.6mm} \textbf{A}_{\hspace{-0.3mm}k} \hspace{-0.3mm} \textbf{C} \right\|_F^2}}{2} \hspace{-0.65mm}-\hspace{-0.55mm} \tfrac{\left\| \textbf{F} \right\|_F^2}{2}  \hspace{-0.65mm}-\hspace{-0.65mm} \tfrac{\left\| \textbf{C}^{\hspace{-0.3mm}H}\hspace{-0.6mm} \textbf{A}_{\hspace{-0.3mm}k} \hspace{-0.3mm} \textbf{C} \right\|_F^2}{2} \hspace{-0.65mm}-\hspace{-0.65mm} \sigma_k^2 t \hspace{-0.65mm}\geq\hspace{-0.5mm} 0, \forall k, \\
\overline{({\rm C7a})}: & \left(x_{n,m}^{\rm Pin} - x_k\right)^2 - z_{n,m,k}^{\rm aff} + \hat{s}_{n,k} \leq 0 , \forall n,m,k,  \\  
\overline{({\rm C7b})}: & \tfrac{1}{\text{Diag} (\textbf{F}_k)_{l(n,m)}} - \hat{s}_{n,k} - x_{n,m,k}^{\rm aff} \leq 0, \forall n,m,k, \\
\overline{({\rm C9b})}: & \theta_{l(n,m),k} -\overline{z_{n,m,k}^{\rm aff}} - \tfrac{2 \pi}{\lambda_g} x_{n,m}^{\rm Pin} \leq 0, \ \forall n, m, k, \\  
\widetilde{\overline{({\rm C11})}}: & \left\|\textbf{X}_k\right\|_* - \left\|\textbf{X}_k^{\rm aff}\right\|_2 \leq 0, \ \forall k, \textbf{X} \in \{\textbf{A}, \textbf{F}\},
\end{align}
where $\widetilde{\overline{({\rm C5})}} \Rightarrow ({\rm C5})$, $\overline{({\rm C7a})} \Rightarrow ({\rm C7a}), \overline{({\rm C7b})} \Rightarrow ({\rm C7b})$, $\overline{({\rm C9b})} \Rightarrow ({\rm C9b})$, and $\widetilde{\overline{({\rm C11})}} \Rightarrow \overline{({\rm C11})}$.
Accordingly, the $(i_3 + 1)$-th iteration of the penalty-based MM method for this subproblem can be expressed as:
\begin{align}\label{eq:sub_formulation3_fin}
  &\underset{\{\,\hat\theta_{i,k},\,\theta_{l(n,m),k},\,t,\,\mathbf F_k,\,\mathbf A_k,\,x_{n,m}^{\rm Pin}\}}%
    {\mathrm{maximize}}\;
    t - \rho \sum_{k=1}^K\sum_{i=2}^{NM}\overline{\Phi}^{(i_3)}_{i,k}
  \nonumber\\
  &\rm{s.t.} \ ({\rm C1}), ({\rm C2}), \widetilde{\overline{({\rm C5})}}, \overline{({\rm C7a})},\overline{({\rm C7b})}, ({\rm C9a}), \overline{({\rm C9b})}, \nonumber \\
  &\textcolor{white}{\rm S. t.} \ ({\rm C10}), \widetilde{\overline{({\rm C11})}}, ({\rm C12}), 
\end{align}
where the resulting problem in~\eqref{eq:sub_formulation3_fin} is convex and can be directly solved using standard convex optimization solvers. This yields a suboptimal solution to the problem~\eqref{eq:sub_formulation3}. 
\begin{figure}[!t]
\centering
\begin{minipage}{0.46\textwidth}
\begin{algorithm}[H] 
\caption{ Overall BCD Algorithm}
\begin{algorithmic}[1]
\State  \textbf{Initialize:} Set the maximum number of iterations $\tau_{\max}$, the convergence tolerance $\tau_{\rm tol} \rightarrow 0$, initialize the index of the previous iteration $\tau = 0$, the optimization variables in $\mathbf{w}^{(\tau)}$, $\textbf{p}_n^{(\tau)}$, $\textbf{x}_n^{\text{Pin}(\tau)}$, and $t^{(\tau)}$.
\State  \textbf{repeat} \hfill \texttt{$\triangleright$ Main Loop: BCD}
\State  \hspace{1em} Obtain $\mathbf{w}^{(\tau+1)}$ with given $\textbf{p}_n^{(\tau)}$ and $\textbf{x}_n^{\text{Pin}(\tau)}$ by solving sub-problem $1$ until convergence;
\State  \hspace{1em} Obtain $\textbf{p}_n^{(\tau+1)}$ with given $\textbf{w}^{(\tau+1)}$ and $\textbf{x}_n^{\text{Pin}(\tau)}$ by solving sub-problem $2$ until convergence;
\State  \hspace{1em} Obtain $\textbf{x}_n^{\text{Pin}(\tau+1)}$ with given $\textbf{w}^{(\tau+1)}$ and $\textbf{p}_n^{(\tau+1)}$ by solving sub-problem $3$ until convergence;
\State  \hspace{1em} Update $t^{(\tau+1)}$ and $\tau = \tau + 1$;
\State  \textbf{until} $\tau = \tau_{\max}$ or $\tfrac{\left|t{(\tau)} - t^{(\tau-1)} \right|}{t{(\tau)}} \leq \tau_{\rm tol}$.
\State  \textbf{return} $\{\textbf{w}^*, \textbf{p}_n^*, \textbf{x}_n^{{\rm Pin}*}, t^*\} = \{\textbf{w}^{(\tau)}, \textbf{p}_n^{(\tau)}, \textbf{x}_n^{{\rm Pin}{(\tau)}}, t^{(\tau)}\}$.
\end{algorithmic}
\end{algorithm}
\end{minipage}
\end{figure}
At each iteration of the BCD algorithm, one block of optimization variables is optimized until convergence to a predefined tolerance, while other blocks remain fixed. We summarize the proposed BCD-based algorithm in \textbf{Algorithm 1}, which is guaranteed to converge to a suboptimal solution of~\eqref{eq:formulation} within polynomial-time computational complexity. The complexity of each iteration of the proposed BCD algorithm is given by $\mathcal{O}\hspace{-0.8mm}\left(\hspace{-0.8mm}I_1\hspace{-0.5mm}(\hspace{-0.5mm}N \hspace{-0.5mm}+ \hspace{-0.5mm}K\hspace{-0.5mm})^3 \hspace{-0.8mm}+\hspace{-0.8mm} I_2\hspace{-0.5mm}\left(\hspace{-0.5mm}NM \hspace{-0.5mm} +\hspace{-0.5mm} K\hspace{-0.5mm}\right)^3 \hspace{-0.8mm}+\hspace{-0.8mm} I_3\left[\hspace{-0.5mm}\left(\hspace{-0.5mm}NM\hspace{-0.5mm}\right)^3\hspace{-0.5mm} + \hspace{-0.5mm}2KM^3\hspace{-0.5mm}\right)\hspace{-0.3mm}\right]$, where $I_1$, $I_2$, and $I_3$ are the numbers of iterations for $3$ subproblems, respectively~\cite{yu2021irs}.

\section{Simulation Results}
\begin{figure}[t]
\centerline{\includegraphics[width=3.35in]{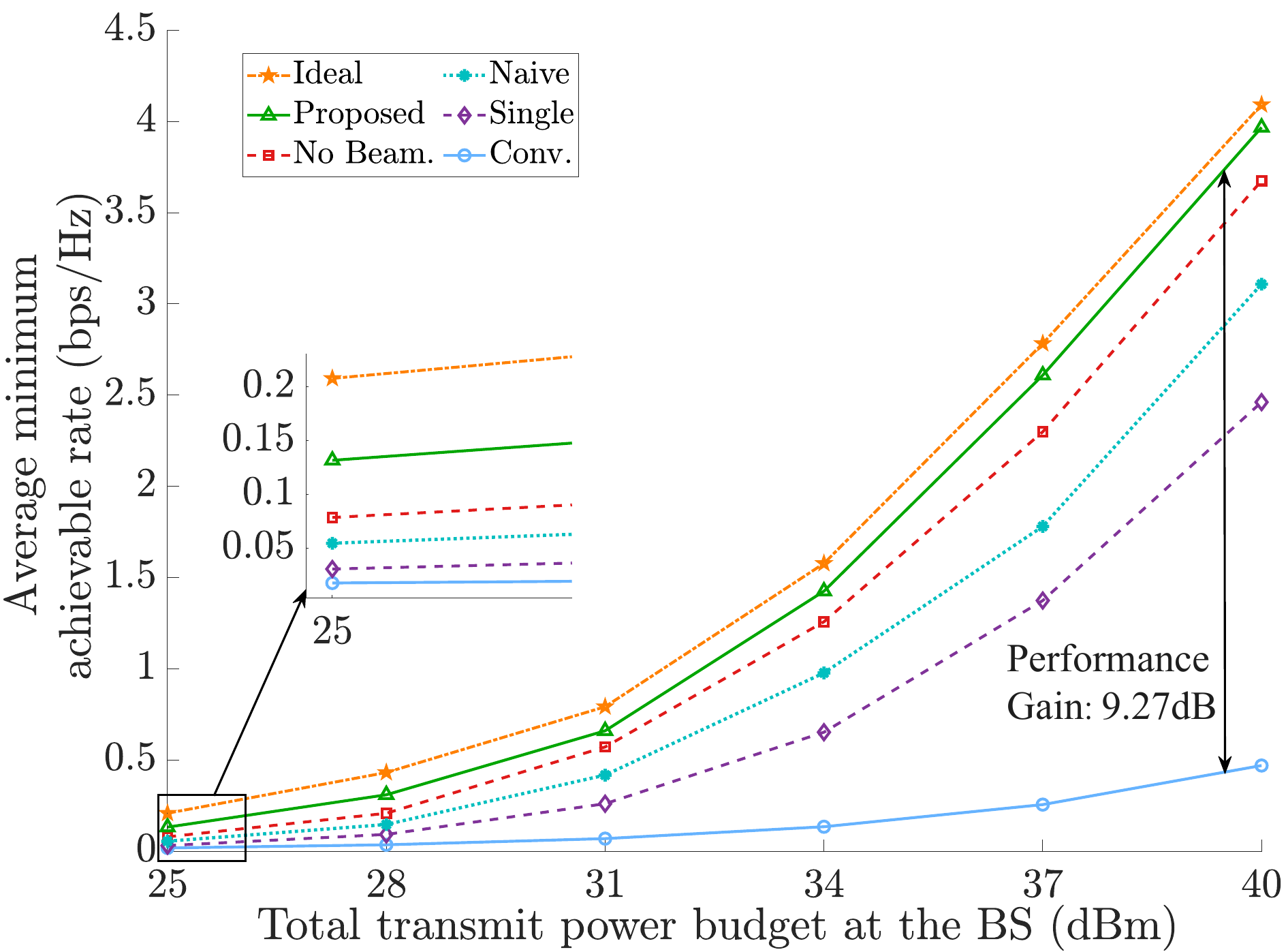}}
\caption{Average minimum achievable rate versus the total transmit power budget of different antenna systems with $N = 2$, $M = 3$, and $K = 5$.}
\label{fig: powerInc}
\vspace{-0.5em}
\end{figure}

This section evaluates the performance of the proposed multi-waveguide multi-PA-assisted broadcast system at $f_c = 28$ GHz via numerical simulations averaged multiple random user location realizations within a $40\,\text{m}\times 40\,\text{m}$ area, based on the configuration depicted in Fig. 1. Dielectric waveguides are fabricated of polytetrafluoroethylene (PTFE)~\cite{suzuki2022pinching} with parameters: $\eta_{\rm eff}=1.42$, $\epsilon_r=2.1$, ${\rm tan}(\delta_e)=2\times10^{-4}$. Simulation-specific parameters ($P_{\max}$, $N$, $M$, and $K$) are detailed in the captions of the respective figures.
We compare the performance of five benchmark schemes against the proposed scheme labeled as ``\textit{Proposed}'' in Fig.~\ref{fig: powerInc}:  1) Ideal PA: Denoted as ``\textit{Ideal}'', which is identical to the proposed approach but assumes zero power attenuation, serving as a theoretical performance upper bound;
2) No beamforming: Denoted as ``\textit{No Beam.}'', which follows the proposed optimization structure but only omits the beamforming design at the BS with assigning equal weight to all waveguides, i.e., $w_n \hspace{-1.5mm} = \hspace{-1.5mm}\tfrac{1}{\sqrt{N}}, \forall n$;
3) Single PA: Denoted as ``\textit{Single}'', which restricts each waveguide to utilize a single PA; 4) Naive baseline: Denoted as ``\textit{Naive}'', which initially optimizes resource allocation ignoring attenuation effects and subsequently recalculates system performance under realistic attenuation condition, if the resulting solution violates constraint (C3), $p_{n,m}=p_{n,m}^{\max}, \forall n,m$;
5) Conventional antenna: Denoted as ``\textit{Conv.}'', this baseline adopts a BS precoder design with antennas fixed at their feed points.

Fig.~\ref{fig: powerInc} demonstrates the minimum achievable rate among all $K$ users as a function of the total transmit power budget $P_{\max}$, comparing the proposed PA-based schemes with conventional antenna schemes. 
The proposed multi-PA-assisted scheme consistently outperforms the conventional antenna scheme, primarily because the optimized placement of multiple PAs can flexibly establish more favorable propagation paths, effectively reducing the large-scale path loss between the BS and users.
As anticipated, the performance gains become more pronounced with increased $P_{\max}$, reaching $9.27$ dB when $P_{\max} = 40$ dBm, reflecting greater flexibility in resource allocation.
Notably, although the ``Single'' benchmark surpasses the ``Conv.'' scheme, it remains inferior to the proposed multi-PA design, as a single PA per waveguide still struggles to effectively mitigate path loss simultaneously for multiple users.
Also, the ``No Beam.'' scheme falls short compared to the proposed approach, as BS beamforming offers additional flexibility to shape signal directions, allowing for precise signal adjustment through the adaptive placement of the PA-induced phase shift.
Besides, the performance of the proposed multi-PA design closely approaches the idealized upper bound, ``Ideal'', in which the latter assumes perfect dielectric waveguides without power attenuation. Furthermore, the ``Naive'' scheme is outperformed by the proposed method, as neglecting practical power attenuation in dielectric waveguides reduces resource allocation accuracy and degrades system performance.
\begin{figure}[t]
    \centering
    \subfigure[$P_{\max} = 30$ dBm.]{
        \includegraphics[width=3.1in]{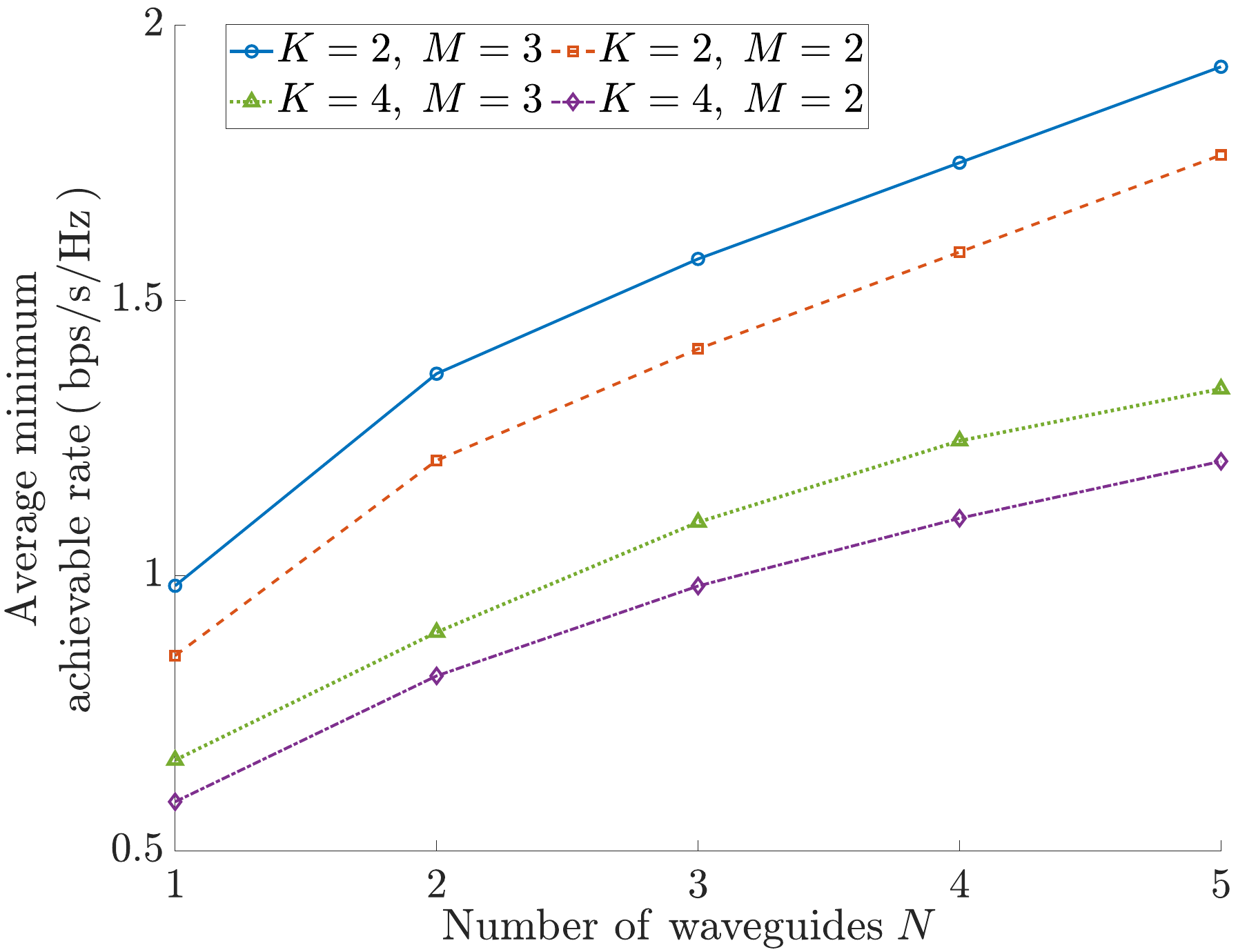}
    }
    \vspace{0.25em}
    \subfigure[$N = 5$, $M = 3$, $K = 4$.]{
        \includegraphics[width=3.1in]{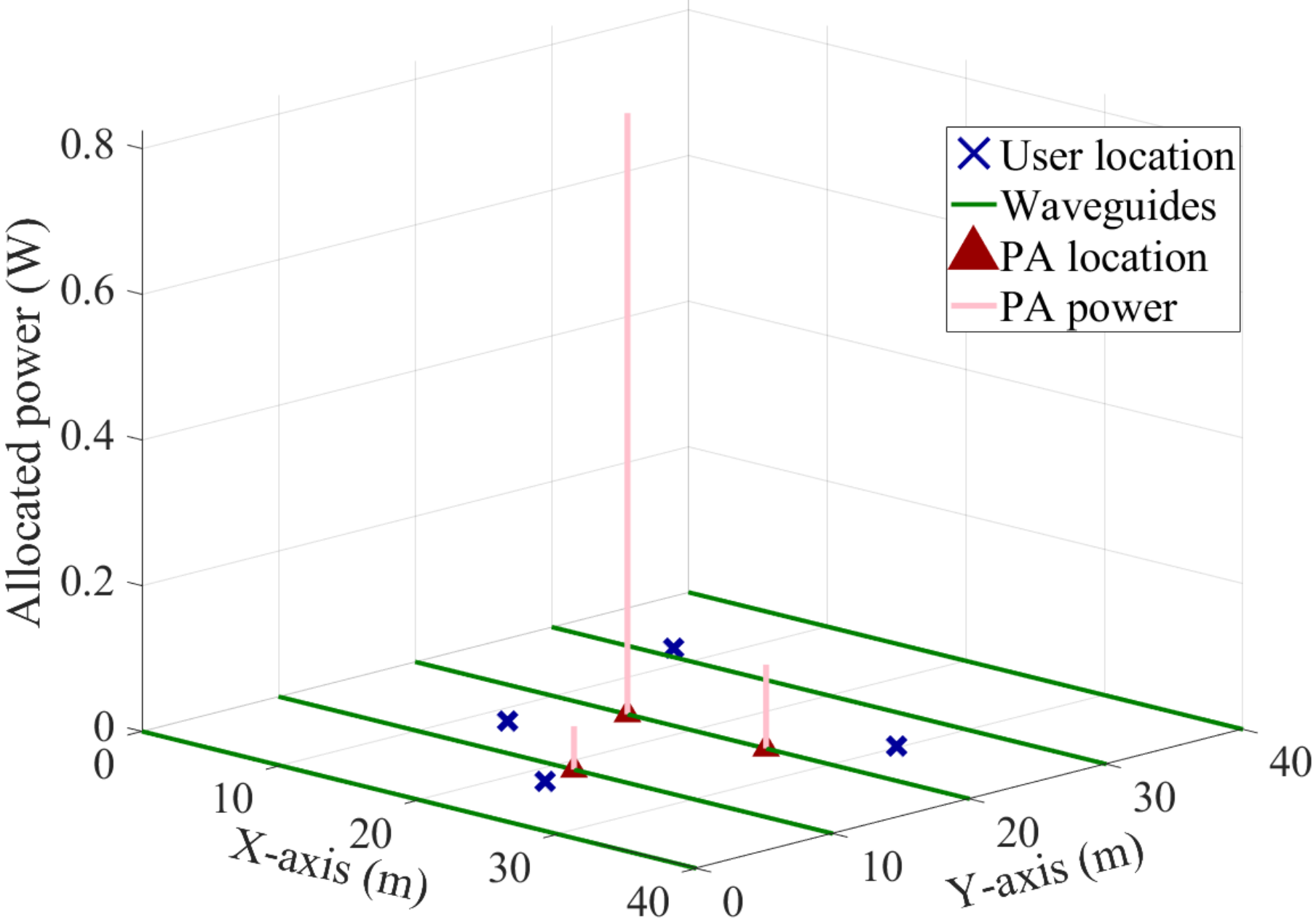}
    }
    \caption{Average minimum achievable rate versus the number of waveguides for different user and antenna settings, with a snapshot example of optimized PA locations and power allocations.}
    \label{fig: usercha}
    \vspace{-0.5em}
\end{figure}

Fig.~\ref{fig: usercha}(a) illustrates the average minimum achievable rate versus the number of waveguides, each with MM PAs, along with a channel realization snapshot in Fig.~\ref{fig: usercha}(b).
From Fig.~\ref{fig: usercha}(a), the average minimum achievable rate improves with increased $N$ and $M$ due to the enhanced spatial DoF for joint beamforming. However, increasing the number of users $K$ reduces performance, as the beamformer must satisfy more constraints, diminishing the available spatial DoF. Consequently, system performance is limited by the user with the weakest channel, whose effective gain decreases as more users are added. Fig.~\ref{fig: usercha}(b) illustrates that out of $15$ available PAs, the algorithm activates only three. Notably, the PA allocated with the highest power is strategically placed near the centroid of the region spanned by all user locations. This placement effectively balances propagation losses within both the dielectric waveguide and wireless channels, enhancing overall system performance and efficiently utilizing transmit power by minimizing energy wasted on distant antennas.

\section{Conclusion}
This paper investigated a multiuser broadcast system assisted by multiple waveguides, each employing multiple PAs, explicitly considering frequency-dependent power attenuation. To maximize the minimum achievable user rate, we formulated a non-convex optimization problem jointly optimizing beamforming precoders, PA ratio allocation, and spatial placement. A BCD-based algorithm using MM and penalty methods was proposed to obtain an effective suboptimal solution. Simulation results demonstrated that the proposed system significantly outperforms conventional antenna systems by up to $9.27$ dB under high-power conditions, revealing that not all PAs are activated due to power and propagation constraints.


\end{document}